\documentclass[showpacs,twocolumn,floatfix,eqsecnum]{revtex4}
\usepackage{graphicx}
\begin{document}
\title{Transition from pure-state to mixed-state entanglement by random scattering}
\author{J. L. van Velsen and C. W. J. Beenakker}
\affiliation{Instituut-Lorentz, Universiteit Leiden, P.O. Box 9506, 2300 RA Leiden, The Netherlands}
\date{March 2004}
\begin{abstract}
We calculate the effect of polarization-dependent scattering by disorder on the degree of polarization-entanglement of two beams of radiation. Multi-mode
detection converts an initially pure state into a mixed state with respect to the polarization degrees of freedom. The degree of
entanglement decays exponentially with the number of detected modes if the scattering mixes the polarization directions and
algebraically if it does not.
\end{abstract}
\maketitle

\section{Introduction}

A pair of photons in the Bell state $\left(|{\rm HV}\rangle + |{\rm VH}\rangle \right)/\sqrt{2}$ can be transported over long
distances with little degradation of the entanglement of their horizontal (${\rm H}$) and vertical (${\rm V}$) polarizations.
Polarization-dependent scattering has little effect on the degree of entanglement, as long as it remains linear (hence describable by a 
scattering matrix) and as long as the photons are detected in a single spatial mode only. This robustness of photon
entanglement was demonstrated dramatically in a recent experiment \cite{Alt02} and theory \cite{Vel03,Mor03} on plasmon-assisted
entanglement transfer.

Polarization-dependent scattering may significantly degrade the entanglement in the case of multi-mode detection. Upon summation over
$N$ spatial modes the initially pure state of the Bell pair is reduced to a mixed state with respect to the polarization
degrees of freedom. This loss of purity diminishes the entanglement --- even if the
two polarization directions are not mixed by the scattering.

The transition from pure-state to mixed-state entanglement will in general depend on the detailed form of the scattering matrix. However,
a universal regime is entered in the case of randomly located scattering centra. This is the regime of applicability of
random-matrix theory \cite{Bee97,Guh98}. As we will show in this paper, the transmission of polarization-entangled
radiation through disordered media reduces the degree of entanglement in a way which, on average, depends only on the number
$N$ of detected modes. (The average refers to an ensemble of disordered media with different random positions of the scatterers.)
The degree of entanglement (as quantified either by the concurrence \cite{Woo98} or by the violation of a Bell inequality
\cite{Bel64,Cla69}) decreases exponentially with $N$ if the disorder randomly mixes the polarization directions. If the 
polarization is conserved, then the decrease is a power law ($\propto N^{-1}$ if both photons are scattered and $\propto 
N^{-1/2}$ if only one photon is scattered).

\begin{figure}
\includegraphics[width=8cm]{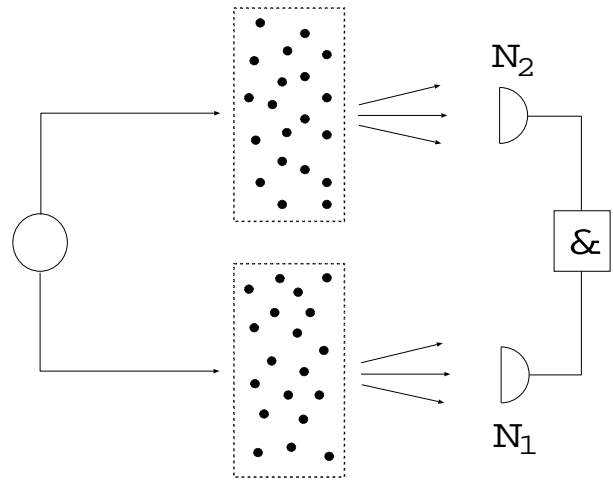}
\caption{
Schematic diagram of the transfer of polarization-entangled radiation through two disordered media. The degree of entanglement of the transmitted radiation is measured by two multi-mode photodetectors ($N_{i}$ modes) in a coincidence circuit. The combination of polarization-dependent scattering and multi-mode detection causes a transition from a pure state to a mixed state in the polarization degrees of freedom, and a resulting decrease of the detected entanglement.
\label{randomscattering}
}
\end{figure}

\section{Formulation of the problem}

We consider two beams of polarization-entangled photons (Bell pairs) that are scattered by two separate disordered media (see Fig.\ \ref{randomscattering}). Two photodetectors in a coincidence circuit measure the degree of entanglement of the transmitted radiation through the violation of a Bell inequality. The scattered Bell pair is in the pure state
\begin{equation}
\Psi_{n\sigma,m\tau}=\frac{1}{\sqrt{2}}\left(u^{+}_{n\sigma}v^{-}_{m\tau}
+u^{-}_{n\sigma}v^{+}_{m\tau}\right).
\end{equation}
The indices $n\in\{1,2,\ldots M_{1}\}$, $m\in\{1,2,\ldots M_{2}\}$ label the transverse spatial modes and the indices $\sigma,\tau\in\{+,-\}$ 
label the horizontal and vertical polarizations. The first pair of indices $n,\sigma$ refers to the first photon and the second pair of indices $m,\tau$ refers to the second photon. The scattering amplitudes $u^{+}_{n\sigma}$, $u^{-}_{n\sigma}$ of the first medium form two orthonormal vectors of length $M_{1}$, and similarly for the second medium.

A subset of $N_{1}$ out of the $M_{1}$ modes are detected in the first detector. We relabel the modes so that $n=1,2,\ldots N_{1}$ are the detected modes. This subset is contained in the four vectors $u^{++}_{n}\equiv u^{+}_{n+}$, $u^{+-}_{n}\equiv u^{+}_{n-}$, $u^{-+}_{n}\equiv u^{-}_{n+}$, $u^{--}_{n}\equiv u^{-}_{n-}$ of length $N_{1}$ each. We write these vectors in bold face, ${\bf u}_{\pm\pm}$, omitting the mode index. Similarly, the second detector detects $N_{2}$ modes, contained in vectors ${\bf v}_{\pm\pm}$. A single or double dot between two pairs of vectors denotes a single or double contraction over the mode indices: ${\bf a}\cdot{\bf b}=\sum_{n}a_{n}b_{n}$,
${\bf a}{\bf b}:{\bf c}{\bf d}=\sum_{n,m}a_{n}b_{m}c_{m}d_{n}$.

The pure state has density matrix $\Psi^{\vphantom{\ast}}_{n\sigma,m\tau}\Psi^{\ast}_{n'\sigma',m'\tau'}$. By tracing over the detected modes the pure state is reduced to a mixed state with respect to the polarization degrees of freedom. The reduced density matrix is $4\times 4$, with elements
\begin{widetext}
\begin{eqnarray}
&&\rho_{\sigma\tau,\sigma'\tau'}=\frac{1}{Z}\bigl({\bf u}_{+\sigma}{\bf
v}_{-\tau}+{\bf u}_{-\sigma}{\bf v}_{+\tau}\bigr):\bigl({\bf
v}^{\ast}_{-\tau'}{\bf u}^{\ast}_{+\sigma'}+{\bf v}^{\ast}_{+\tau'}{\bf
u}^{\ast}_{-\sigma'}\bigr),\label{rhomixed}\\
&&Z=\sum_{\sigma,\tau}\bigl({\bf u}_{+\sigma}{\bf v}_{-\tau}+{\bf
u}_{-\sigma}{\bf v}_{+\tau}\bigr):\bigl({\bf v}^{\ast}_{-\tau}{\bf
u}^{\ast}_{+\sigma}+{\bf v}^{\ast}_{+\tau}{\bf u}^{\ast}_{-\sigma}\bigr). \label{Zdef}
\end{eqnarray}
The complex numbers that enter into the density matrix are conveniently grouped into a pair of Hermitian positive definite matrices $a$ and $b$, with elements $a_{\sigma\tau,\sigma'\tau'}={\bf u}^{\vphantom{\ast}}_{\sigma\tau}\cdot{\bf u}^{\ast}_{\sigma'\tau'}$,  $b_{\sigma\tau,\sigma'\tau'}={\bf v}^{\vphantom{\ast}}_{\sigma\tau}\cdot{\bf v}^{\ast}_{\sigma'\tau'}$. One has
\begin{equation}
Z\rho_{\sigma\tau,\sigma'\tau'}=a_{+\sigma,+\sigma'}b_{-\tau,-\tau'}+a_{-\sigma,-\sigma'}b_{+\tau,+\tau'}+a_{-\sigma,+\sigma'}b_{+\tau,-\tau'}+a_{+\sigma,-\sigma'}b_{-\tau,+\tau'}. \label{rhoarelation}
\end{equation}
\end{widetext}

The degree of entanglement of the mixed state with $4\times 4$ density matrix $\rho$ is quantified by the concurrence ${\cal C}$, given by \cite{Woo98}
\begin{equation}
{\cal C}=\rm{max}\left\{0,\sqrt{\lambda_{1}}-\sqrt{\lambda_{2}}-\sqrt{\lambda_{3}}-\sqrt{\lambda_{4}}\right\}. \label{Cdef}
\end{equation}
The $\lambda_{i}$'s are the eigenvalues of the matrix product
\[
\rho\cdot (\sigma_{y}\otimes\sigma_{y})\cdot\rho^{*}\cdot (\sigma_{y}\otimes\sigma_{y}),
\]
in the order $\lambda_{1}\geq\lambda_{2}\geq\lambda_{3}\geq\lambda_{4}$, with $\sigma_{y}$ a Pauli matrix. The concurrence ranges from 0 (no entanglement) to 1 (maximal entanglement).

In a typical experiment \cite{Alt02}, the photodetectors can not measure ${\cal C}$ directly, but instead infer the degree of entanglement through the maximal violation of the Bell-CHSH (Clauser-Horne-Shimony-Holt) inequality \cite{Bel64,Cla69}. The maximal value ${\cal E}$ of the Bell-CHSH parameter for an arbitrary mixed state was analyzed in Refs.\ \cite{Hor95,Ver02}. For a pure state with concurrence ${\cal C}$ one has simply ${\cal E}=2\sqrt{1+{\cal C}^{2}}$ \cite{Gis91}. For a mixed state there is no one-to-one relation between ${\cal E}$ and ${\cal C}$. Depending on the density matrix, ${\cal E}$ can take on values between $2{\cal C}\sqrt{2}$ and $2\sqrt{1+{\cal C}^{2}}$, so ${\cal E}>2$ implies ${\cal C}>0$ but not the other way around. The general formula
\begin{equation}
{\cal E}=2\sqrt{u_{1}+u_{2}}\label{Emaxformula}
\end{equation}
for the dependence of ${\cal E}$ on $\rho$ involves the two largest eigenvalues $u_{1},u_{2}$ of the real symmetric $3\times 3$ matrix $R^{\rm T}R$ constructed from  $R_{kl}={\rm Tr}\,\rho\,\sigma_{k}\otimes\sigma_{l}$. Here $\sigma_{1},\sigma_{2},\sigma_{3}$ refer to the three Pauli matrices $\sigma_{x},\sigma_{y}$, $\sigma_{z}$, respectively.

We will calculate both the true concurrence ${\cal C}$ and the pseudo-concurrence
\begin{equation}
{\cal C}'\equiv \sqrt{\max\left(0,{\cal E}^{2}/4-1\right)}\leq {\cal C} \label{Epseudo}
\end{equation}
inferred from the Bell inequality violation.

As a special case we will also consider what happens if only one of the two beams is scattered. The other beam reaches the photodetector without changing its mode or polarization, so we may set $v^{\pm}_{m\sigma}=\delta_{m,1}\delta_{\sigma,\pm}$. This implies $b_{\sigma\tau,\sigma'\tau'}=\delta_{\sigma,\tau}\delta_{\sigma',\tau'}$, hence
\begin{equation}
Z\rho_{\sigma\tau,\sigma'\tau'}=a_{\bar{\tau}\sigma,\bar{\tau}'\sigma'},\label{rho1medium}
\end{equation}
where we have defined $\bar{\tau}=-\tau$. The normalization is now given simply by $Z=\sum_{\sigma,\tau}a_{\sigma\tau,\sigma\tau}$.

\section{Random-matrix theory}
\label{RMT}

For a statistical description we use results from the random-matrix theory (RMT) of scattering by disordered media \cite{Bee97,Guh98}. According to that theory, the real and imaginary parts of the complex scattering amplitudes $u^{\sigma\tau}_{n}$ are statistically distributed as independent random variables with the same Gaussian distribution of zero mean. The variance of the Gaussian drops out of the density matrix; we fix it at 1. The assumption of independent variables ignores the orthonormality constraint of the vectors $u$, which is justified if $N_{1}\ll M_{1}$. Similarly, for $N_{2}\ll M_{2}$ the real and imaginary parts of $v^{\sigma\tau}_{n}$ have independent Gaussian distributions with zero mean and a variance which we may set at 1.

The reduced density matrix of the mixed state depends on the two independent random matrices $a$ and $b$, according to Eq.\ (\ref{rhoarelation}). The matrix elements are not independent. We calculate the joint probability distribution of the matrix elements, using  the following result from RMT \cite{Ver94}: Let $W$ be a rectangular matrix of dimension $p\times(k+p)$, filled with complex numbers with distribution
\begin{equation}
P(\{W_{nm}\})\propto \exp\left(-c\,{\rm Tr}\,WW^{\dagger}\right),\;\;c>0.
\label{propW}
\end{equation}
Then the square matrix $H=WW^{\dagger}$ (of dimension $p\times p$) has the Laguerre distribution
\begin{equation}
P(\{H_{nm}\})\propto ({\rm Det}\,H)^{k} \exp(-c\,{\rm Tr}\,H).
\end{equation}
Note that $H$ is Hermitian and positive definite, so its eigenvalues $h_{n}$ ($n=1,2,\ldots p$) are real positive numbers. Their joint distribution is that of the Laguerre unitary ensemble,
\begin{equation}
P(\{h_{n}\})\propto \prod_{n} h_{n}^{k}\,e^{-c h_{n}}\prod_{i<j}(h_{i}-h_{j})^{2}.
\end{equation}
The factor $(h_{i}-h_{j})^{2}$ is the Jacobian of the transformation from complex matrix elements to real eigenvalues. The eigenvectors of $H$ form a unitary matrix $U$ which is uniformly distributed in the unitary group.

To apply this to the matrix $a$ we set $c=1/2$, $p=4$, $k=N_{1}-4$. We first assume that $N_{1}\geq 4$, to ensure that $k\geq 0$. Then
\begin{eqnarray}
&&P(\{a_{\sigma\tau,\sigma'\tau'}\})\propto ({\rm Det}\,a)^{N_{1}-4} \exp\left(-{\textstyle\frac{1}{2}}\,{\rm Tr}\,a\right), \label{Pasigmatau}\\
&&P(\{a_{n}\})\propto \prod_{n} a_{n}^{N_{1}-4}e^{-a_{n}/2}\prod_{i<j}(a_{i}-a_{j})^{2}, \label{Pan1}
\end{eqnarray}
where $a_{1},a_{2},a_{3},a_{4}$ are the real positive eigenvalues of $a$. The $4\times 4$ matrix $U$ of eigenvectors of $a$ is uniformly distributed in the unitary group. If $N_{1}=1,2,3$ we set $c=1/2$, $p=N_{1}$, $k=4-N_{1}$. The matrix $a$ has $4-N_{1}$ eigenvalues equal to $0$. The $N_{1}$ non-zero eigenvalues have distribution
\begin{equation}
P(\{a_{n}\})\propto \prod_{n} a_{n}^{4-N_{1}}e^{-a_{n}/2}\prod_{i<j}(a_{i}-a_{j})^{2}. \label{Pan2}
\end{equation}
The distribution of the matrix elements $b_{\sigma\tau,\sigma'\tau'}$ and of the eigenvalues $b_{n}$ is obtained upon replacement of $N_{1}$ by $N_{2}$ in Eqs.\  (\ref{Pasigmatau}), (\ref{Pan1}), and (\ref{Pan2}).

\section{Asymptotic analysis}
\label{Asymptotics}

We wish to average the concurrence (\ref{Cdef}) and pseudo-concurrence (\ref{Epseudo}) with the RMT distribution of 
Sec.\ \ref{RMT}. The result depends only on the number of detected modes $N_{1},N_{2}$ in the two photodetectors. 
Microscopic details of the scattering media become irrelevant once we assume random scattering. 
The averages $\langle{\cal C}\rangle$, $\langle{\cal C}'\rangle$ can be calculated by numerical integration \cite{Num04}. 
Before presenting these results, we analyze the asymptotic behavior for $N_{i}\gg 1$ analytically. 
We assume for simplicity that $N_{1}=N_{2}\equiv N$.

It is convenient to scale the eigenvalues as
\begin{equation}
a_{n}=2N(1+\alpha_{n}),\;\; b_{n}=2N(1+\beta_{n}).\label{rescaling}
\end{equation}
The distribution of the $\alpha_{n}$'s and $\beta_{n}$'s takes the same form
\begin{equation}
P(\{\alpha_{n}\})\propto\exp\left(-N\sum_{n=1}^{4}[\alpha_{n}-\ln(1+\alpha_{n})]+{\cal O}(1)\right),\label{Palpha}
\end{equation}
where ${\cal O}(1)$ denotes $N$-independent terms. 
The bulk of the distribution (\ref{Palpha}) lies in the region $\sum_{n}\alpha_{n}^{2}\alt 1/N\ll 1$, localized at the origin. 
Outside of this region the distribution decays exponentially $\propto \exp[-Nf(\{\alpha_{n}\})]$, 
with 
\begin{equation}
f(\{\alpha_{n}\})=\sum_{n=1}^{4} [\alpha_{n}-\ln(1+\alpha_{n})]. 
\end{equation}

The concurrence ${\cal C}$ and pseudo-concurrence ${\cal C}'$ depend on the rescaled eigenvalues $\alpha_{n},\beta_{n}$ and also on the pair of $4\times 4$ unitary matrices $U,V$ of eigenvectors of $a$ and $b$. Both quantities are {\em independent\/} of $N$, because the scale factor $N$ in Eq.\ (\ref{rescaling}) drops out of the density matrix (\ref{rhoarelation}) upon normalization. 

The two quantities ${\cal C}$ and ${\cal C}'$ are identically zero when the $\alpha_{n}$'s and $\beta_{n}$'s are all $\ll 1$ in absolute value. 
For a nonzero value one has to go deep into the tail of the eigenvalue distribution. 
The average of ${\cal C}$ is dominated by the 
``optimal fluctuation'' $\alpha_{n}^{\rm opt}$, $\beta_{n}^{\rm opt}$, $U^{\rm opt}$, $V^{\rm opt}$ of eigenvalues and eigenvectors, 
which minimizes $f(\{\alpha_{n}\})+f(\{\beta_{n}\})$ in the region ${\cal C}>0$. The decay
\begin{equation}
\langle{\cal C}\rangle\simeq\exp\left(-N[f(\{\alpha_{n}^{\rm opt}\})+f(\{\beta_{n}^{\rm opt}\})]\right)
\equiv e^{-AN} \label{Cdecay}
\end{equation}
of the average concurrence is exponential in $N$, with a coefficient $A$ of order unity determined by the optimal fluctuation. 
The average $\langle{\cal C}'\rangle \simeq e^{-BN}$ also decays exponentially with $N$, but with a different coefficient $B$ 
in the exponent.
The numbers $A$ and $B$ can be calculated analytically for the case that only one of the two beams is scattered.

\begin{figure}[h!]
\includegraphics[width=8cm]{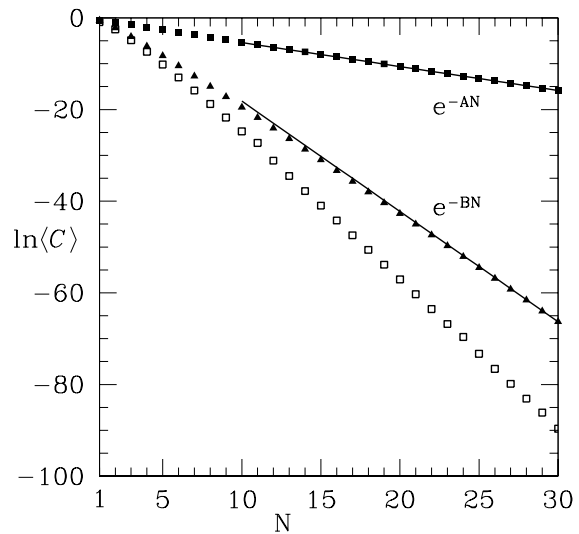}
\caption{
Average concurrence $\langle{\cal C}\rangle$ (squares) and pseudo-concurrence $\langle{\cal C}'\rangle$ (triangles) as a function of the 
number $N$ of detected modes. Closed symbols are for the case that only one of the two beams is scattered and open symbols for the case that
both beams are scattered. The decay of $\langle{\cal C}'\rangle$ in the latter case could not be determined accurately enough and is therefore
omitted from the plot. The solid lines are the analytically obtained exponential decays, with constants $A=3\ln 3 -4\ln 2$ and 
$B=\ln (11+5\sqrt{5}) -\ln 2$, cf. Eqs. (\ref{Adecay}) and (\ref{Bdecay}). 
\label{Cfig}}
\end{figure}

Scattering of a single beam corresponds to a density matrix $\rho$ which is directly given by the matrix $a$, cf. Eq. (\ref{rho1medium}).
To find $A$, we therefore need to minimize $f(\{\alpha_{n}\})$ 
over the eigenvalues and eigenvectors of $a$ with the constraint ${\cal C} > 0$,
\begin{equation}
A=\min_{\{\alpha_{n}\},U} \left\{f(\{\alpha_{n}\})|\,{\cal C}\bigl(\rho(\{\alpha_{n}\},U)\bigr)>0\right\}
\label{singlescatmin}.
\end{equation}
The minimum can be found with the help of the following result \cite{Ver01}: 
The concurrence ${\cal C}(\rho)$ of the two-qubit density matrix $\rho$, with fixed eigenvalues
$\Lambda_{1} \geq \Lambda_{2} \geq \Lambda_{3} \geq \Lambda_{4}$ but arbitrary
eigenvectors, is maximized upon unitary transformation by
\begin{equation}
\max_{\Omega}\, {\cal C}(\Omega\,\rho\,\Omega^{\dagger})=\max\left\{0,\Lambda_{1}-\Lambda_{3}-2\sqrt{\Lambda_{2}\Lambda_{4}}\right\}.
\end{equation}
(The matrix $\Omega$ varies over all $4 \times 4$ unitary matrices.)
With this knowledge, Eq. (\ref{singlescatmin}) reduces to  
\begin{equation}
A = \min_{\{\alpha_{n}\}} \big\{f(\{\alpha_{n}\})|   
\alpha_{1}-\alpha_{3}-2\sqrt{(1+\alpha_{2})(1+\alpha_{4})}>0\big\},
\end{equation}
where we have ordered $\alpha_{1} \geq \alpha_{2} \geq \alpha_{3} \geq \alpha_{4}$.
This yields for the optimal fluctuation 
$\alpha_{1}^{\rm opt}=1$, $\alpha_{2}^{\rm opt}=\alpha_{3}^{\rm opt}=\alpha_{4}^{\rm opt}=-1/3$ and
\begin{equation}
A=3\ln 3 -4\ln 2 = 0.523.
\label{Adecay}
\end{equation}
The asymptotic decay $\langle{\cal C}\rangle \propto e^{-A N}$ is in
good agreement with a numerical calculation for finite $N$, see Fig.\ \ref{Cfig}.

The asymptotic decay of the average pseudo-concurrence $\langle{\cal C}'\rangle$ for a single scattered beam can be found in a similar way,
using the result \cite{Ver02} 
\begin{widetext}
\begin{equation}
\max_{\Omega}\, {\cal C}'(\Omega\,\rho\,\Omega^{\dagger})=  
\sqrt{\max\left\{0,2(\Lambda_{1}-\Lambda_{4})^{2}+2(\Lambda_{2}-\Lambda_{3})^{2}-
(\Lambda_{1}+\Lambda_{2}+\Lambda_{3}+\Lambda_{4})^{2}\right\} }. 
\end{equation}
To obtain the optimal fluctuation we have to solve
\begin{equation}
B=\min_{\{\alpha_{n}\}} \big\{f(\{\alpha_{n}\})| 
2(\alpha_{1}-\alpha_{4})^{2}+2(\alpha_{2}-\alpha_{3})^{2}- 
(4+\alpha_{1}+\alpha_{2}+\alpha_{3}+\alpha_{4})^{2} > 0 \big\},
\end{equation}
which gives 
\begin{equation}
\alpha_{1}^{\rm opt}=\frac{1}{2}(-1+2\sqrt{2}+\sqrt{5}), \; \; 
\alpha_{2}^{\rm opt}=\alpha_{3}^{\rm opt}=\frac{1}{2}(1-\sqrt{5}), \; \;  
\alpha_{4}^{\rm opt}=\frac{1}{2}(-1-2\sqrt{2}+\sqrt{5}),  
\end{equation}
\end{widetext}
hence
\begin{equation}
B=\ln (11+5\sqrt{5}) -\ln 2 = 2.406.
\label{Bdecay}
\end{equation}
The decay $\langle{\cal C}'\rangle \propto e^{-B N}$ is again in good agreement with the numerical results 
for finite $N$ (Fig.\ \ref{Cfig}).   

If both beams are scattered, a calculation of the optimal fluctuation is more complicated because the
eigenvalues $\{\alpha_{n}\}$, $\{\beta_{n}\}$ and the eigenvectors $U$, $V$ get mixed in the density matrix (\ref{rhoarelation}).
The numerics of Fig.\ \ref{Cfig} gives $\langle{\cal C}\rangle \propto e^{-3.3 N}$ for the asymptotic decay of the concurrence. 
The averaged pseudo-concurrence for two-beam scattering could not be determined accurately enough to extract a reliable value for
the decay constant.    

\section{Comparison with the case of polarization-conserving scattering}

If the scatterers are translationally invariant in one direction, then the two polarizations are not mixed by the
scattering. Such scatterers have been realized as parallel glass fibers \cite{Mon91}. One polarization 
corresponds to the
electric field parallel to the scatterers ({\rm TE} polarization), the other to parallel magnetic field 
({\rm TM} polarization). The boundary condition differs for the two polarizations (Dirichlet for {\rm TE} and 
Neumann for {\rm TM}), so the scattering amplitudes ${\bf u}_{++}$, ${\bf v}_{++}$, ${\bf u}_{--}$, ${\bf v}_{--}$
that conserve the polarization can still be considered to be independent random numbers. The amplitudes that couple
different polarizations vanish: ${\bf u}_{+-}$, ${\bf v}_{+-}$, ${\bf u}_{-+}$, ${\bf v}_{-+}$ are all zero.

The reduced density matrix (\ref{rhoarelation}) simplifies to
\begin{equation} 
Z\rho_{\sigma\tau,\sigma'\tau'}=
\delta_{\sigma\bar{\tau}}\delta_{\sigma'\bar{\tau}'}a_{\sigma\sigma,\sigma'\sigma'}b_{\tau\tau,\tau'\tau'},
\label{rhoconservedpol}
\end{equation}
with $\bar{\tau}=-\tau$, $\bar{\tau}'=-\tau'$.
We will abbreviate $A_{\sigma\tau} \equiv a_{\sigma\sigma,\tau\tau}$, $B_{\sigma\tau} \equiv b_{\sigma\sigma,\tau\tau}$.
The concurrence ${\cal C}$ and pseudo-concurrence ${\cal C}'$ are calculated from Eqs. (\ref{Cdef}) and 
(\ref{Epseudo}), with the result
\begin{equation}
{\cal C}={\cal C}'=\frac{2|A_{+-}||B_{+-}|}{A_{++}B_{--}+A_{--}B_{++}}.
\label{Cconservedpol}
\end{equation}     

\begin{figure}
\includegraphics[width=8cm]{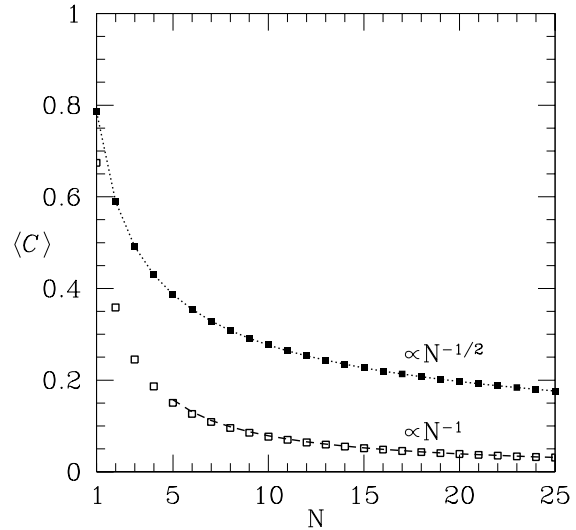}
\caption{
Average concurrence $\langle{\cal C}\rangle$ as a function of the number $N$ of detected modes, for the case of 
polarization-conserving scattering of both beams (open squares) and one beam (closed squares). The
data points are the result of a numerical average.
The dashed line is the asymptotic result (\ref{Cconservedpolasym}) and the dotted line is the analytical result 
(\ref{avCconservedpolonescat}). The pseudo-concurrence ${\cal C}'$ is identical to ${\cal C}$ for polarization-conserving
scattering.
\label{Cconservedpolfig}}
\end{figure}

It is again our objective to calculate  $\langle{\cal C}\rangle$ for the case $N_{1}=N_{2}=N$. 
The distribution of the matrices $A$ and $B$ follows
by substituting $N_{1}-4 \rightarrow N-2$ in Eq. (\ref{Pasigmatau}):
\begin{equation}
P(\{A_{\sigma\tau}\})\propto ({\rm Det}\,A)^{N-2} \exp\left(-{\textstyle\frac{1}{2}}\,{\rm Tr}\,A\right).
\label{PAsigmatau}
\end{equation}
The average over this distribution was done numerically, see Fig.\ \ref{Cconservedpolfig}.
For large $N$ we may perform the following asymptotic analysis. 

We scale the matrices $A$ and $B$ as  
\begin{equation}
A=2N(\openone + {\cal A}), \; \; B=2N(\openone + {\cal B}).
\end{equation}
In the limit $N \rightarrow \infty$ the Hermitian matrices ${\cal A}$ and ${\cal B}$ 
have the Gaussian distribution
\begin{equation}
P(\{{\cal A}_{\sigma\tau}\})\propto
e^{-\frac{1}{2}N{\rm Tr {\cal A}{\cal A}^{\dagger}}}.
\label{GUE2}
\end{equation}
(The same distribution holds for ${\cal B}$.)
In contrast to the analysis in Sec.\ \ref{Asymptotics} the concurrence   
does not vanish in the bulk of the distribution. The average of Eq. (\ref{Cconservedpol}) with 
distribution (\ref{GUE2}) yields the algebraic decay 
\begin{equation}  
\langle{\cal C}\rangle=\frac{\pi}{4}\frac{1}{N}, \; \; N \gg 1,
\label{Cconservedpolasym}
\end{equation}
in good agreement with the numerical calculation for finite $N$ (Fig.\ \ref{Cconservedpolfig}).

A completely analytical calculation for any $N$ can be done in the case that only one of the beams
is scattered. In that case $B_{\sigma\tau}=1$ and the concurrence reduces to
\begin{equation}
{\cal C}=\frac{2|A_{+-}|}{A_{++}+A_{--}}.
\label{Cconservedpolonescat}
\end{equation}      
Averaging Eq. (\ref{Cconservedpolonescat}) over the Laguerre distribution (\ref{PAsigmatau}) gives 
\begin{equation}
\langle{\mathcal C}\rangle=\frac{\sqrt{\pi}}{2}\frac{\Gamma(N+1/2)}{\Gamma(N+1)}.
\label{avCconservedpolonescat}
\end{equation}
For large $N$, the average concurrence 
(\ref{avCconservedpolonescat}) falls off as 
\begin{equation}
\langle{\mathcal C}\rangle=\frac{\sqrt{\pi}}{2}\frac{1}{\sqrt{N}}, \; \; N \gg 1. 
\end{equation}
This case is also included in Fig.\ \ref{Cconservedpolfig}.

\section{Conclusion}
In summary, we have applied the method of random-matrix theory ({\rm RMT}) to the problem of entanglement
transfer through a random medium. {\rm RMT} has been used before to study {\em production} of entanglement
\cite{Fur98,Mil99,Zyc01,Ban02,Zni03,Sco03,Jac03,Bee03}. Here we have studied the {\em loss} of entanglement in the transition 
from a pure state to a mixed state.

A common feature of all these theories is that the results are universal, independent of microscopic details.
In our problem the decay of the degree of entanglement depends on the number of detected modes but not on 
microscopic parameters such as the scattering mean free path.

The origin of this universality is the central limit theorem: The complex scattering amplitude from one mode
in the source to one mode in the detector is the sum over a large number of complex partial amplitudes, 
corresponding to different sequences of multiple scattering. The probability distribution of the sum becomes a
Gaussian with zero mean (because the random phases of the partial amplitudes average out to zero). The variance
of the Gaussian will depend on the mean free path, but it drops out upon normalization of the reduced density
matrix. The applicability of the central limit theorem only requires that the separation of source and detector 
is large compared to the scattering mean free path, to ensure a large number of terms in the sum over partial
amplitudes.

The degree of entanglement (as quantified by the concurrence or violation of the Bell inequality) then depends only
on the number $N$ of detected modes. We have identified two qualitatively different types of decay. The decay is
exponential $\propto e^{-cN}$ if the scattering mixes spatial modes as well as polarization directions.
The coefficient $c$ depends on which measure of entanglement one uses (concurrence or violation of
Bell inequality) and it also depends on whether both photons in the Bell pair are scattered or only one of them is. 
For this latter case of single-beam scattering, the coefficients $c$ are $3\ln3 - 4\ln2$ (concurrence) and $\ln(11+5\sqrt{5})-\ln2$ 
(pseudo-concurrence). The decay is
algebraic $\propto N^{-p}$ if the scattering preserves the polarization. The power $p$ is 1 if both photons are 
scattered and $1/2$ if only one of them is. Polarization-conserving scattering is special;
it would require translational invariance of the scatterers in one direction. The generic decay is therefore
exponential.

Finally, we remark that the results presented here apply not only to scattering by disorder, but also to
scattering by a cavity with a chaotic phase space. An experimental search for entanglement loss by chaotic 
scattering has been reported  by Woerdman {\em et al.} \cite{Woe03}. 

\acknowledgments
This work was supported by the ``Stichting voor Fundamenteel Onderzoek der Materie'' (FOM), by the \linebreak
``Nederlandse organisatie voor Wetenschappelijk Onderzoek'' (NWO), and by the U.S. Army Research Office (Grant No.
DAAD 19-02-0086).


\begin{thebibliography}{99}
\bibitem{Num04} 
The results of Fig.\ \ref{Cfig} are obtained by
a combination of adaptive integration over the simplex of eigenvalues 
$\{a_{n}\}$ (and $\{b_{n}\}$) and a stochastic average 
over the unitary matrix $U$ (and $V$). We define the scaled eigenvalues $\{\tilde{a}_{n}\}$ by $a_{2}=a_{1}\tilde{a}_{2}$,
$a_{3}=a_{1}\tilde{a}_{2}\tilde{a}_{3}$, $a_{4}=a_{1}\tilde{a}_{2}\tilde{a}_{3}\tilde{a}_{4}$, and similarly for $\{\tilde{b}_{n}\}$. 
The eigenvalues $a_{1}$ (and $b_{1}$) are then integrated out
analytically. This can be done since $\rho$ (and hence ${\cal C}$ and ${\cal C}'$) is left invariant upon scaling 
$\{a_{n}\}$ and $\{b_{n}\}$. 
In the case that one beam is scattered, we averaged over 4000 random unitary matrices $U$ and
used a maximum of 6000 integration points in the cube of scaled eigenvalues ($0 \leq \tilde{a}_{n} \leq 1, \, n=2,3,4$).
For scattering of both beams, an average over 1000 pairs of $U$ and $V$ was taken with a maximum of $12\cdot 10^{5}$ points
in the 6-dimensional space of combined cubes.
\end{thebibliography}
\end{document}